\pdfoutput=1

\documentclass[12pt,a4paper]{article}

\usepackage{ifthen} 
\newboolean{pdflatex}
\setboolean{pdflatex}{true} 

\newboolean{articletitles}
\setboolean{articletitles}{true} 

\newboolean{uprightparticles}
\setboolean{uprightparticles}{false} 

\newboolean{inbibliography}
\setboolean{inbibliography}{false} 

\usepackage{microtype}
\usepackage{lineno}  
\usepackage{xspace} 
\usepackage{caption} 

\usepackage{graphicx}  
\usepackage{color}
\usepackage{colortbl}
\graphicspath{{./figs/}} 

\usepackage{amsmath} 
\usepackage{amssymb}
\usepackage{amsfonts}
\usepackage{upgreek} 

\usepackage{booktabs}

\usepackage{tabularx}
\usepackage{rotating}

\newcommand*\patchAmsMathEnvironmentForLineno[1]{%
\expandafter\let\csname old#1\expandafter\endcsname\csname #1\endcsname
\expandafter\let\csname oldend#1\expandafter\endcsname\csname
end#1\endcsname
 \renewenvironment{#1}%
   {\linenomath\csname old#1\endcsname}%
   {\csname oldend#1\endcsname\endlinenomath}%
}
\newcommand*\patchBothAmsMathEnvironmentsForLineno[1]{%
  \patchAmsMathEnvironmentForLineno{#1}%
  \patchAmsMathEnvironmentForLineno{#1*}%
}
\AtBeginDocument{%
\patchBothAmsMathEnvironmentsForLineno{equation}%
\patchBothAmsMathEnvironmentsForLineno{align}%
\patchBothAmsMathEnvironmentsForLineno{flalign}%
\patchBothAmsMathEnvironmentsForLineno{alignat}%
\patchBothAmsMathEnvironmentsForLineno{gather}%
\patchBothAmsMathEnvironmentsForLineno{multline}%
}

\usepackage{hyperref}    
\usepackage[all]{hypcap} 

\def\lhcb {\mbox{LHCb}\xspace}

\ifthenelse{\boolean{uprightparticles}}%
{

 \def\Pmu         {\ensuremath{\upmu}\xspace}

 \def\Ppi         {\ensuremath{\uppi}\xspace}

 \def\Ppsi        {\ensuremath{\uppsi}\xspace}

 \def\PDelta      {\ensuremath{\Delta}\xspace}                 
 \def\PXi      {\ensuremath{\Xi}\xspace}                 
 \def\PLambda      {\ensuremath{\Lambda}\xspace}                 
 \def\PSigma      {\ensuremath{\Sigma}\xspace}                 
 \def\POmega      {\ensuremath{\Omega}\xspace}                 
 \def\PUpsilon      {\ensuremath{\Upsilon}\xspace}                 
 
 \def\PB      {\ensuremath{\mathrm{B}}\xspace}                 
                  
 \def\PD      {\ensuremath{\mathrm{D}}\xspace}

 \def\PJ      {\ensuremath{\mathrm{J}}\xspace}                 
 \def\PK      {\ensuremath{\mathrm{K}}\xspace}

 \def\Pb      {\ensuremath{\mathrm{b}}\xspace}                 
 \def\Pc      {\ensuremath{\mathrm{c}}\xspace}                 
 \def\Pd      {\ensuremath{\mathrm{d}}\xspace}

 \def\Pi      {\ensuremath{\mathrm{i}}\xspace}

 \def\Pp      {\ensuremath{\mathrm{p}}\xspace}

 \def\Ps      {\ensuremath{\mathrm{s}}\xspace}                 
                  
 \def\Pu      {\ensuremath{\mathrm{u}}\xspace}

}
{

 \def\Pmu         {\ensuremath{\mu}\xspace}

 \def\Ppi         {\ensuremath{\pi}\xspace}

 \def\Ppsi        {\ensuremath{\psi}\xspace}                 
                  
 \mathchardef\PDelta="7101
 \mathchardef\PXi="7104
 \mathchardef\PLambda="7103
 \mathchardef\PSigma="7106
 \mathchardef\POmega="710A
 \mathchardef\PUpsilon="7107
                  
 \def\PB      {\ensuremath{B}\xspace}                 
                  
 \def\PD      {\ensuremath{D}\xspace}

 \def\PJ      {\ensuremath{J}\xspace}                 
 \def\PK      {\ensuremath{K}\xspace}

 \def\Pb      {\ensuremath{b}\xspace}                 
 \def\Pc      {\ensuremath{c}\xspace}                 
 \def\Pd      {\ensuremath{d}\xspace}

 \def\Pi      {\ensuremath{i}\xspace}

 \def\Pp      {\ensuremath{p}\xspace}

 \def\Ps      {\ensuremath{s}\xspace}                 
                  
 \def\Pu      {\ensuremath{u}\xspace}

}

\def\mun        {\ensuremath{\Pmu^-}\xspace} 

\def\uquark    {\ensuremath{\Pu}\xspace}

\def\dquark    {\ensuremath{\Pd}\xspace}

\def\squark    {\ensuremath{\Ps}\xspace}

\def\cquark    {\ensuremath{\Pc}\xspace}

\def\bquark    {\ensuremath{\Pb}\xspace}

\def\pion  {\ensuremath{\Ppi}\xspace}

\def\pip   {\ensuremath{\pion^+}\xspace}
\def\pim   {\ensuremath{\pion^-}\xspace}

\def\kaon  {\ensuremath{\PK}\xspace}
  \def\Kbar  {\kern 0.2em\overline{\kern -0.2em \PK}{}\xspace}

\def\Kp    {\ensuremath{\kaon^+}\xspace}
\def\Km    {\ensuremath{\kaon^-}\xspace}

\def\KS    {\ensuremath{\kaon^0_{\rm\scriptscriptstyle S}}\xspace}

  \def\Dbar    {\kern 0.2em\overline{\kern -0.2em \PD}{}\xspace}
\def\D       {\ensuremath{\PD}\xspace}

\def\Dzb     {\ensuremath{\Dbar^0}\xspace}
\def\Dp      {\ensuremath{\D^+}\xspace}

\def\Dstar   {\ensuremath{\D^*}\xspace}

\def\Dstarp  {\ensuremath{\D^{*+}}\xspace}
\def\Dstarm  {\ensuremath{\D^{*-}}\xspace}

\def\Dsp     {\ensuremath{\D^+_\squark}\xspace}

\def\B       {\ensuremath{\PB}\xspace}
\def\Bbar    {\ensuremath{\kern 0.18em\overline{\kern -0.18em \PB}{}}\xspace}
\def\Bb      {\ensuremath{\Bbar}\xspace}

\def\Bd      {\ensuremath{\B^0}\xspace}
\def\Bs      {\ensuremath{\B^0_\squark}\xspace}
\def\Bsb     {\ensuremath{\Bbar^0_\squark}\xspace}
\def\Bdb     {\ensuremath{\Bbar^0}\xspace}

\def\jpsi     {\ensuremath{{\PJ\mskip -3mu/\mskip -2mu\Ppsi\mskip 2mu}}\xspace}

  \def\Y#1S{\ensuremath{\PUpsilon{(#1S)}}\xspace}

\def\proton      {\ensuremath{\Pp}\xspace}

\def\Lz {\ensuremath{\PLambda}\xspace}
\def\Lbar {\ensuremath{\kern 0.1em\overline{\kern -0.1em\PLambda}}\xspace}

\def\Lb      {\ensuremath{\Lz^0_\bquark}\xspace}

\def\Lc      {\ensuremath{\Lz^+_\cquark}\xspace}

\def\BF         {{\ensuremath{\cal B}\xspace}}

\def\BR         {\BF}

\def\to                 {\ensuremath{\rightarrow}\xspace}

\def\CP                {\ensuremath{C\!P}\xspace}

\def\AT#1     {\ensuremath{A_{\mathrm{T}}^{#1}}\xspace}           

\def\C#1      {\ensuremath{\mathcal{C}_{#1}}\xspace}                       
\def\Cp#1     {\ensuremath{\mathcal{C}_{#1}^{'}}\xspace}                    
\def\Ceff#1   {\ensuremath{\mathcal{C}_{#1}^{\mathrm{(eff)}}}\xspace}        
\def\Cpeff#1  {\ensuremath{\mathcal{C}_{#1}^{'\mathrm{(eff)}}}\xspace}       
\def\Ope#1    {\ensuremath{\mathcal{O}_{#1}}\xspace}                       
\def\Opep#1   {\ensuremath{\mathcal{O}_{#1}^{'}}\xspace}                    

\newcommand{\tev}{\ifthenelse{\boolean{inbibliography}}{\ensuremath{~T\kern -0.05em eV}\xspace}{\ensuremath{\mathrm{\,Te\kern -0.1em V}}\xspace}}
\newcommand{\gev}{\ensuremath{\mathrm{\,Ge\kern -0.1em V}}\xspace}
\newcommand{\mev}{\ensuremath{\mathrm{\,Me\kern -0.1em V}}\xspace}
\newcommand{\kev}{\ensuremath{\mathrm{\,ke\kern -0.1em V}}\xspace}
\newcommand{\ev}{\ensuremath{\mathrm{\,e\kern -0.1em V}}\xspace}
\newcommand{\gevc}{\ensuremath{{\mathrm{\,Ge\kern -0.1em V\!/}c}}\xspace}
\newcommand{\mevc}{\ensuremath{{\mathrm{\,Me\kern -0.1em V\!/}c}}\xspace}
\newcommand{\gevcc}{\ensuremath{{\mathrm{\,Ge\kern -0.1em V\!/}c^2}}\xspace}
\newcommand{\gevgevcccc}{\ensuremath{{\mathrm{\,Ge\kern -0.1em V^2\!/}c^4}}\xspace}
\newcommand{\mevcc}{\ensuremath{{\mathrm{\,Me\kern -0.1em V\!/}c^2}}\xspace}

\def\mum  {\ensuremath{{\,\upmu\rm m}}\xspace}

\def\invfb   {\ensuremath{\mbox{\,fb}^{-1}}\xspace}

\def\fs   {\ensuremath{\rm \,fs}\xspace}

\newcommand{\chisq}{\ensuremath{\chi^2}\xspace}
\newcommand{\chisqndf}{\ensuremath{\chi^2/\mathrm{ndf}}\xspace}
\newcommand{\chisqip}{\ensuremath{\chi^2_{\rm IP}}\xspace}

\def\gsim{{~\raise.15em\hbox{$>$}\kern-.85em
          \lower.35em\hbox{$\sim$}~}\xspace}
\def\lsim{{~\raise.15em\hbox{$<$}\kern-.85em
          \lower.35em\hbox{$\sim$}~}\xspace}

\def\sPlot{\mbox{\em sPlot}\xspace}

\def\pt         {\mbox{$p_{\rm T}$}\xspace}

\def\evtgen     {\mbox{\textsc{EvtGen}}\xspace}

\def\geant      {\mbox{\textsc{Geant4}}\xspace}

\def\photos     {\mbox{\textsc{Photos}}\xspace}

\def\pythia     {\mbox{\textsc{Pythia}}\xspace}

\def\tell1  {TELL1\xspace}
\def\ukl1   {UKL1\xspace}

\newcommand{\ie}{\mbox{\itshape i.e.}\xspace}

\usepackage{cite} 
\usepackage{mciteplus}

\usepackage{longtable} 

\begin{document}

\renewcommand{\thefootnote}{\fnsymbol{footnote}}
\setcounter{footnote}{1}

\def\flfd      {\ensuremath{f_{\Lb}/f_\dquark}\xspace}
\def\fsfd      {\ensuremath{f_{\squark}/f_\dquark}\xspace}
\def\fl      {\ensuremath{f_{\Lb}}\xspace}
\def\fd      {\ensuremath{f_{\dquark}}\xspace}
\def\fs      {\ensuremath{f_{\squark}}\xspace}

\def\DLLp	{\ensuremath{\textrm{DLL}(p-\pi)}\xspace}
\def\DLLK	{\ensuremath{\textrm{DLL}(K-\pi)}\xspace}
\def\DLLpK	{\ensuremath{\textrm{DLL}(p-K)}\xspace}
\def\flfufd      {\ensuremath{f_{\Lb}/(f_\uquark + f_\dquark)}\xspace}
\def\fsfufd      {\ensuremath{f_{\squark}/(f_\uquark + f_\dquark)}\xspace}
\def\fu      {\ensuremath{f_{\uquark}}\xspace}

\def\Hb      	{\ensuremath{H_\bquark}\xspace}
\def\Hc      	{\ensuremath{H_\cquark}\xspace}
\def\Deltapp     {\ensuremath{\Delta^{++}(1232)}\xspace}
\def\Lambdastar     {\ensuremath{\Lambda(1520)}\xspace}
\def\Sg         {\ensuremath{\PSigma_c^+} \xspace}

\def\pKpi     {\ensuremath{\Pp \Km \pip}\xspace}
\def\KPiPi	{\ensuremath{\Km \pip \pip}\xspace}

\def\BsDsPi     {\ensuremath{\Bsb \to \Dsp \pim}\xspace}
\def\BdDPi     {\ensuremath{\Bdb \to \Dp\pim}\xspace}
\def\BdDMuNu     {\ensuremath{\Bdb \to \Dp \mun \bar{\nu}}\xspace}
\def\BdDK     {\ensuremath{\Bdb \to \Dp \Km}\xspace}
\def\LbLcPi	{\ensuremath{\Lb \to \Lc \pim}\xspace}
\def\LbLcMuNu	{\ensuremath{\Lb \to \Lc \mun \bar{\nu}}\xspace}
\def\BsDsK     {\ensuremath{\Bsb \to \Dsp \Km}\xspace}
\def\LcPi	{\ensuremath{\Lc \pim}\xspace}
\def\LcPiWS	{\ensuremath{\Lc \pip}\xspace}
\def\LbLcK	{\ensuremath{\Lb \to \Lc \Km}\xspace}
\def\LcK	{\ensuremath{\Lc \Km}\xspace}
\def\LcpKpi	{\ensuremath{\Lc \to \Pp \Km \pip}\xspace}
\def\DKPiPi	{\ensuremath{\Dp \to \Km \pip \pip}\xspace}
\def\LbSgPi	{\ensuremath{\Lb \to \PSigma_c^+ \pim}\xspace}
\def\LbLcRho	{\ensuremath{\Lb \to \Lc \rho^-}\xspace}
\def\BdDstPi    {\ensuremath{\Bdb \to \Dstarp \pim}\xspace}
\def\BdDRho     {\ensuremath{\Bdb \to \Dp \rho^-}\xspace}
\def\DsKKPi	{\ensuremath{\Dsp \to \Km \Kp \pip}\xspace}
\def\LbLcMuX     {\ensuremath{\Lb \to \Lc \mun X}\xspace}
\def\BdDMuX     {\ensuremath{\Bdb \to \Dp \mun X}\xspace}
\def\BsDsMuX     {\ensuremath{\Bsb \to \Dsp \mun X}\xspace}
\def\LbLcpKsPi	{\ensuremath{\Lb \to \Lc (\to \Pp \KS) \pim}\xspace}

\def\Lbst      {\ensuremath{\Lz^{*0}_\bquark}\xspace}
\def\Sbst {\ensuremath{\PSigma_\bquark^{(*)}}\xspace}

\def\LbLcPifull	  {\ensuremath{\Lb \to \Lc (\to \Pp \Km \pip) \pim}\xspace}
\def\BdDPifull    {\ensuremath{\Bdb \to \Dp (\to \Km \pip \pip) \pim}\xspace}

\newcommand{\cgev}{\ensuremath{{c}\mathrm{/ Ge\kern -0.1em V}}\xspace}

\newcommand{\BRLbResultRound}{
  \ensuremath{ \Big( 4.30 \pm 0.03 \,\, ^{+0.12}_{-0.11} \pm 0.26 \pm 0.21 \Big) \times 10^{-3}}
}


\begin{titlepage}
\pagenumbering{roman}

\vspace*{-1.5cm}
\centerline{\large EUROPEAN ORGANIZATION FOR NUCLEAR RESEARCH (CERN)}
\vspace*{1.5cm}
\hspace*{-0.5cm}
\begin{tabular*}{\linewidth}{lc@{\extracolsep{\fill}}r}
\ifthenelse{\boolean{pdflatex}}
{\vspace*{-2.7cm}\mbox{\!\!\!\includegraphics[width=.14\textwidth]{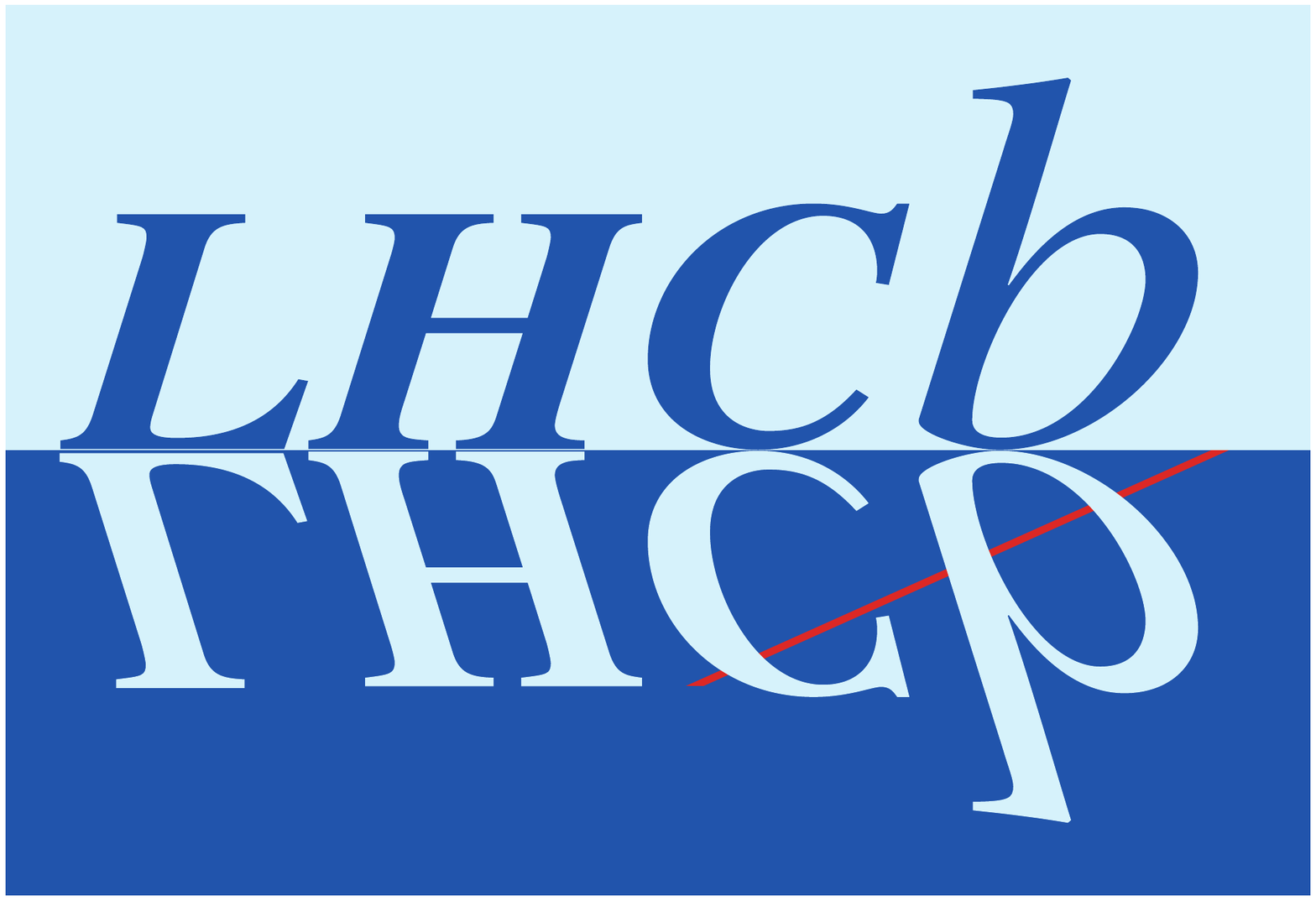}} & &}%
{\vspace*{-1.2cm}\mbox{\!\!\!\includegraphics[width=.12\textwidth]{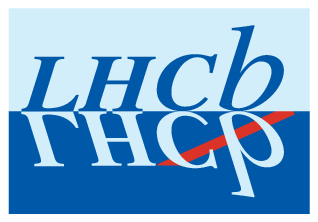}} & &}%
\\
 & &      CERN-PH-EP-2014-106 \\  
 & &      LHCb-PAPER-2014-004 \\  
 & & 27 May 2014 \\ 
 & & \\
\end{tabular*}

\vspace*{2.0cm}

{\bf\boldmath\huge
\begin{center}
  Study of the kinematic dependences of \Lb production in $pp$ collisions and a measurement of the \LbLcPi branching fraction
\end{center}
}

\vspace*{0.6cm}

\begin{center}
The LHCb collaboration\footnote{Authors are listed on the following pages.}
\end{center}


\begin{abstract}
  \noindent
The kinematic dependences of the relative production rates, \flfd, of \Lb baryons and \Bd mesons 
are measured using \LbLcPi and \BdDPi decays.  
The measurements use proton-proton collision data, corresponding to an integrated luminosity of 1~fb$^{-1}$ at 
a centre-of-mass energy of 7~TeV, recorded in the forward region with the LHCb experiment.
The relative production rates are observed to depend on the transverse momentum, \pt, and pseudorapidity, $\eta$, of the beauty hadron,
in the studied kinematic region $1.5 < \pt < 40$~\gevc and $2 < \eta < 5$.
Using a previous \lhcb measurement of \flfd in semileptonic decays, the branching fraction \mbox{$\BF(\LbLcPi) = \BRLbResultRound$} 
is obtained, where the first uncertainty is statistical, the second is systematic, 
the third is from the previous \lhcb measurement of \flfd and the fourth is due to the \BdDPi branching fraction.
This is the most precise measurement of a \Lb branching fraction to date.
\end{abstract}

\vspace*{0.0cm}

\begin{center}
  Submitted to JHEP
\end{center}

\vspace{\fill}

{\footnotesize 
\centerline{\copyright~CERN on behalf of the \lhcb collaboration, license \href{http://creativecommons.org/licenses/by/3.0/}{CC-BY-3.0}.}}
\vspace*{2mm}

\end{titlepage}


\newpage
\setcounter{page}{2}
\mbox{~}
\newpage

\centerline{\large\bf LHCb collaboration}
\begin{flushleft}
\small
R.~Aaij$^{41}$, 
B.~Adeva$^{37}$, 
M.~Adinolfi$^{46}$, 
A.~Affolder$^{52}$, 
Z.~Ajaltouni$^{5}$, 
J.~Albrecht$^{9}$, 
F.~Alessio$^{38}$, 
M.~Alexander$^{51}$, 
S.~Ali$^{41}$, 
G.~Alkhazov$^{30}$
% ...

{\footnotesize \it
$ ^{1}$Centro Brasileiro de Pesquisas F\'{i}sicas (CBPF), Rio de Janeiro, Brazil\\
% ...
$ ^{5}$Clermont Universit\'{e}, Universit\'{e} Blaise Pascal, CNRS/IN2P3, LPC, Clermont-Ferrand, France\\
% ...
$ ^{9}$Fakult\"{a}t Physik, Technische Universit\"{a}t Dortmund, Dortmund, Germany\\
% ...
$ ^{30}$Petersburg Nuclear Physics Institute (PNPI), Gatchina, Russia\\
% ...
$ ^{37}$Universidad de Santiago de Compostela, Santiago de Compostela, Spain\\
$ ^{38}$European Organization for Nuclear Research (CERN), Geneva, Switzerland\\
% ...
$ ^{41}$Nikhef National Institute for Subatomic Physics, Amsterdam, The Netherlands\\
% ...
$ ^{46}$H.H. Wills Physics Laboratory, University of Bristol, Bristol, United Kingdom\\
% ...
$ ^{51}$School of Physics and Astronomy, University of Glasgow, Glasgow, United Kingdom\\
$ ^{52}$Oliver Lodge Laboratory, University of Liverpool, Liverpool, United Kingdom\\
% ...
}
\end{flushleft}


\renewcommand{\thefootnote}{\arabic{footnote}}
\setcounter{footnote}{0}



\pagestyle{plain} 
\setcounter{page}{1}
\pagenumbering{arabic}


%

\newpage
\section{Introduction}
\label{sec:Introduction}
Measurements of beauty hadron production in high-energy proton-proton ($pp$) collisions
provide valuable information on fragmentation and hadronisation 
within the framework of quantum chromodynamics~\cite{Berezhnoy:2013cda}. 
The study of beauty baryon decays also provides an additional channel for investigating \CP violation~\cite{Dunietz:1992ti}.
While significant progress has been made in the understanding of the production and decay properties of beauty mesons, 
knowledge of beauty baryons is limited.

The relative production rates of beauty hadrons are described by the fragmentation fractions $f_u$, $f_d$, $f_s$, $f_c$ and $f_{\Lb}$, 
which describe the probability that a $b$ quark fragments into a $B_q$~meson (where $q=u,d,s,c$) or a \Lb~baryon, respectively,
and depend on the kinematic properties of the $b$ quark.
Strange $b$ baryons are less abundantly produced~\cite{LHCb-PAPER-2014-010} and are neglected here.
Measurements of ground state $b$~hadrons produced at the $pp$ interaction point also include decay products of excited $b$~hadrons.
In the case of $B$ mesons, such excited states include $B^{*}$ and $B^{**}$ mesons, while \Lb baryons can be 
produced via decays of \Lbst or \Sbst baryons.

Knowledge of the relative production rate of \Lb~baryons is necessary to measure absolute \Lb branching fractions. 
The measurement of the branching fraction of the \LbLcPi decay  reported in this paper improves the determination of any \Lb branching 
fraction measured relative to the \LbLcPi decay.
The inclusion of charge conjugate processes is implied throughout this paper.
The average branching fraction and production ratios are measured.

Previous measurements of \flfd have been made in $e^+ e^-$ collisions at LEP~\cite{HFAG}, $p\overline{p}$ 
collisions at CDF~\cite{Aaltonen:2008zd,Aaltonen:2008eu} and $pp$ collisions at LHCb~\cite{LHCb-PAPER-2011-018}. 
The value of \flfd measured at LEP differs significantly from the values measured at the hadron colliders, indicating
a strong dependence of \flfd on the kinematic properties of the $b$ quark.

The LHCb analysis~\cite{LHCb-PAPER-2011-018} was based on semileptonic $\Lb \rightarrow \Lc\mu^{-}\bar{\nu} X$ and
$\Bb \rightarrow \D \mu^{-} \bar{\nu} X$ decays, where the $B$ meson is charged or neutral, and
$X$ represents possible additional decay products of the $b$ hadron that are not included in the candidate reconstruction.
Near equality of the inclusive semileptonic decay width of all $b$~hadrons was assumed.
The analysis measured \flfufd, which can be converted into \flfd under the assumption of isospin symmetry, \ie $\fu = \fd$.
A clear dependence of \flfd on the transverse momentum \pt of the $\Lc\mu^-$ and $\D\mu^-$ pairs was observed. 
A CMS analysis~\cite{CMS_Lb} using $\Lb \rightarrow \jpsi \Lz$ decays also found that the cross-section for \Lb~baryons fell faster 
with \pt than the $b$-meson cross-sections.

The present paper uses a data sample, corresponding to an integrated luminosity of 1\invfb of $pp$ collision data at a
centre-of-mass energy of 7 TeV, collected with the LHCb detector.
This is a substantially increased data sample compared to that in Ref.~\cite{LHCb-PAPER-2011-018}.
The analysis aims to clarify the extent and characteristics of the \pt dependence of \flfd.
Moreover, the dependence of \flfd on the pseudorapidity $\eta$, defined in terms of the polar angle $\theta$ with respect to the beam 
direction as $-\ln(\tan{\theta/2})$, is studied for the first time.
The analysis covers the fiducial region $1.5 < \pt < 40$~\gevc and $2 < \eta < 5$.

The hadronic decays \LbLcPi and \BdDPi are used, with the charm hadrons reconstructed using the decay modes \LcpKpi and \DKPiPi, respectively.
The data sample and the selection of \BdDPi decays are identical to those used in Ref.~\cite{LHCb-PAPER-2012-037}.
Although a precise measurement of the absolute value of \flfd is not possible with these decays, since the \LbLcPi branching fraction is
poorly known~\cite{PDG2012}, they can be used to measure the dependence of \flfd on the $b$-hadron kinematic properties to high precision. 
This is achieved by measuring the efficiency-corrected yield ratio $\mathcal{R}$ in bins of \pt or $\eta$ of the beauty hadron 
\begin{equation}
\label{eq:DefnOfR}
 \mathcal{R} (x) \equiv \frac{N_{\LbLcPi} (x) }{N_{\BdDPi} (x) } \times \frac{\varepsilon_{ \BdDPi} (x)}{\varepsilon_{\LbLcPi} (x)},
\end{equation}
where $N$ is the event yield, $\epsilon$ is the total reconstruction and selection efficiency, and $x$ denotes \pt or $\eta$.
The quantity $\mathcal{R}$ is related to \flfd through
\begin{eqnarray}
\label{eq:DefnOfS}
\frac{\fl}{\fd} (x) &=& \frac{\BR(\BdDPi)}{\BR(\LbLcPi)} \times \frac{\BR(\DKPiPi)}{\BR(\LcpKpi)}  \times  
                               \mathcal{R} (x) \nonumber \\
                    & & \nonumber \\
                           &\equiv& \mathcal{S} \times  \mathcal{R} (x),
\end{eqnarray}
where $\mathcal{S}$ is a constant scale factor.

Since the value of \flfd in a given bin of \pt or $\eta$ is independent of the decay mode of the $b$~hadron, the values of 
$\flfd(\pt)$ from the semileptonic analysis~\cite{LHCb-PAPER-2011-018} can be compared to the measurement of $\mathcal{R}(\pt)$, 
which allows for the extraction of the value of $\mathcal{S}$. 
The branching fraction $\BR(\LbLcPi)$ can then be readily obtained using Eq.~\eqref{eq:DefnOfS}.
Notably, the dependence on $\BR(\LcpKpi)$ cancels when extracting $\BR(\LbLcPi)$ in this way, because this branching fraction 
also enters in the semileptonic measurement of \flfd. Furthermore, the branching fractions $\BR(\BdDPi)$~\cite{PDG2012} and 
$\BR(\DKPiPi)$~\cite{Dobbs:2007ab} are well known, leading to a precise determination of $\BR(\LbLcPi)$.

The dependence of the semileptonic \flfd measurement on $\BR(\Lb \rightarrow \Lc\mu^{-}\bar{\nu} X)$, and the assumption of near 
equality of the inclusive semileptonic decay width of all $b$~hadrons, implies that the measurement of $\BR(\LbLcPi)$ from the current 
paper cannot be used to normalise existing measurements of $\BR(\Lb \rightarrow \Lc\mu^{-}\bar{\nu} X)$~\cite{PDG2012}.

\section{Detector and simulation}
\label{sec:Detector}

The \lhcb detector~\cite{Alves:2008zz} is a single-arm forward
spectrometer covering the \mbox{pseudorapidity} range $2<\eta <5$,
designed for the study of particles containing \bquark or \cquark
quarks. The detector includes a high-precision tracking system
consisting of a silicon-strip vertex detector surrounding the $pp$
interaction region, a large-area silicon-strip detector located
upstream of a dipole magnet with a bending power of about
$4{\rm\,Tm}$, and three stations of silicon-strip detectors and straw
drift tubes placed downstream.
The combined tracking system provides a momentum measurement with
relative uncertainty that varies from 0.4\% at 5\gevc to 0.6\% at 100\gevc,
and impact parameter resolution of 20\mum for
tracks with large \pt. 
Different types of charged hadrons are distinguished by information
from two ring-imaging Cherenkov detectors. Photon, electron and
hadron candidates are identified by a calorimeter system consisting of
scintillating-pad and preshower detectors, an electromagnetic
calorimeter and a hadronic calorimeter. Muons are identified by a system 
composed of alternating layers of iron and multiwire proportional chambers.

The trigger consists of a hardware stage, based on information from the calorimeter and muon
systems, followed by a software stage, which applies a full event reconstruction.
The events used in this analysis are selected at the hardware stage by requiring a cluster in
the calorimeters with transverse energy greater than $3.6\gev$.
The software trigger requires a two-, three- or four-track secondary vertex~(SV) with a large sum 
of the \pt of the particles and a significant displacement from the primary $pp$ interaction 
vertices~(PVs). At least one charged particle should have $\pt > 1.7\gevc$ and \chisqip 
with respect to any PV greater than 16, where \chisqip is defined as the
difference in fit \chisq of a given PV reconstructed with and without the considered track.
A multivariate algorithm is used for the identification of SVs consistent with 
the decay of a \bquark hadron.

Simulated collision events are used to estimate the efficiency of the reconstruction and
selection steps for signal as well as background $b$-hadron decay modes.
In the simulation, $pp$ collisions are generated using \pythia~\cite{Sjostrand:2006za}
with a specific \lhcb configuration~\cite{LHCb-PROC-2010-056}. Decays of hadronic particles
are described by \evtgen~\cite{Lange:2001uf}, in which final-state radiation is generated using 
\photos~\cite{Golonka:2005pn}. The interaction of the generated particles with the detector and 
its response are implemented using the \geant toolkit~\cite{Allison:2006ve, *Agostinelli:2002hh} 
as described in Ref.~\cite{LHCb-PROC-2011-006}.

\section{Event selection}
\label{sec:EvtSel}

Since the \LbLcPifull and \BdDPifull decays have the same topology, 
the criteria used to select them are chosen to be similar.  
This minimises the systematic uncertainty on the ratio of the selection efficiencies.
Following the trigger selection, a preselection is applied using the reconstructed masses, decay times and
vertex qualities of the $b$-hadron and $c$-hadron candidates.  
Further separation between signal and background is achieved using a boosted decision tree (BDT)~\cite{Breiman}.  
The BDT is trained and tested on a sample of \BsDsPi events from the same data set as the signal events.
This sample of events is not used elsewhere in the analysis. 
For the signal, a weighted data sample based on the \sPlot technique~\cite{Pivk:2004ty} is used. 
A training sample representative of combinatorial background is selected 
from \Bs candidates with mass greater than 5445~\mevcc.
The variables with the most discriminating power are found to be the \chisqip of the $b$-hadron candidate 
with respect to the PV, the \pt of the final-state particles, and the angle between the $b$-hadron momentum vector 
and the vector connecting its production and decay vertices. 
In events with multiple PVs, the $b$ hadron is associated to the PV giving the smallest \chisqip.
  
The BDT requirement is chosen to maximise the signal yield divided by the square root of the sum of the signal and
background yields. It rejects approximately 84\% of the combinatorial background events while retaining 
approximately 84\% of the signal events.
The $\Dp$ ($\Lc$) candidates are identified by requiring the invariant mass under the \KPiPi (\pKpi) hypothesis
to fall within the range 1844--1890~(2265--2305)~\mevcc.
The mass resolution of the charm hadrons is approximately 6~\mevcc.

The ratio of selection efficiencies is evaluated using simulated events.  
The \linebreak \DKPiPi decay is generated using the known Dalitz structure~\cite{Bonvicini:2008jw},  
while the \LcpKpi decay is generated using a combination of non-resonant and resonant decay modes with 
proportions according to Ref.~\cite{Aitala:1999uq}.
Interference effects in the \Lc decay are not taken into account. 
Consistency checks, using a phase-space only model for the \LcpKpi decay, show negligible differences in
the relative efficiencies.
The distributions of the input variables to the BDT are compared in data and simulation.
Good agreement is observed for most variables. 
The largest deviation is seen for quantities related to the track quality. 
The simulated events are reweighted so that the distributions of these quantities reproduce the distributions in data. 

The final stage of the event selection applies particle identification (PID) criteria on all tracks, 
based on the differences in the natural logarithm of the likelihood between the 
pion, kaon and proton hypotheses~\cite{LHCb-DP-2012-003}.  
The PID performance as a function of the \pt and $\eta$ of the charged particle is estimated from data.
This is performed using calibration samples, selected using only kinematic criteria, and consisting of 
approximately 27 million $\Dstarm \to \Dzb(\Kp\pim)\pim$ decays for kaons and pions, 
and 13 million $\Lz \to \proton \pim$ decays for protons. 
The size of the proton calibration sample is small at high \pt of the proton and does not allow 
a reliable estimate of the efficiency of the proton PID requirement in this kinematic region. 
Hence, proton PID criteria are only applied to candidates restricted to 
a kinematic region in proton momentum and pseudorapidity corresponding to low-\pt protons.
Outside of this region, no PID criteria are imposed on the proton.

\begin{figure}[!b]
    \begin{picture}(500,180)(0,0)
     \put(-5,1){\includegraphics[scale=0.4]{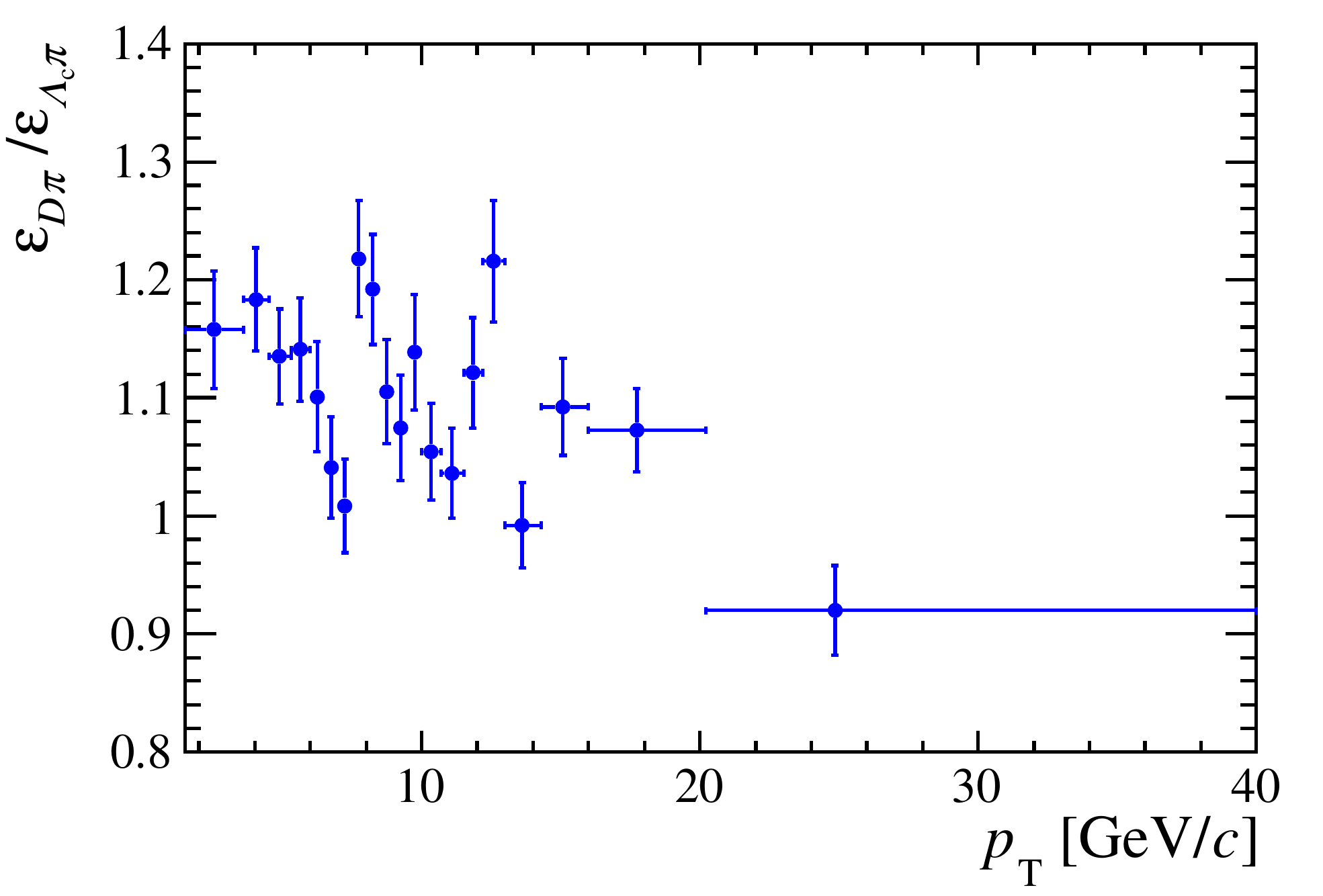}}
     \put(230,1){\includegraphics[scale=0.4]{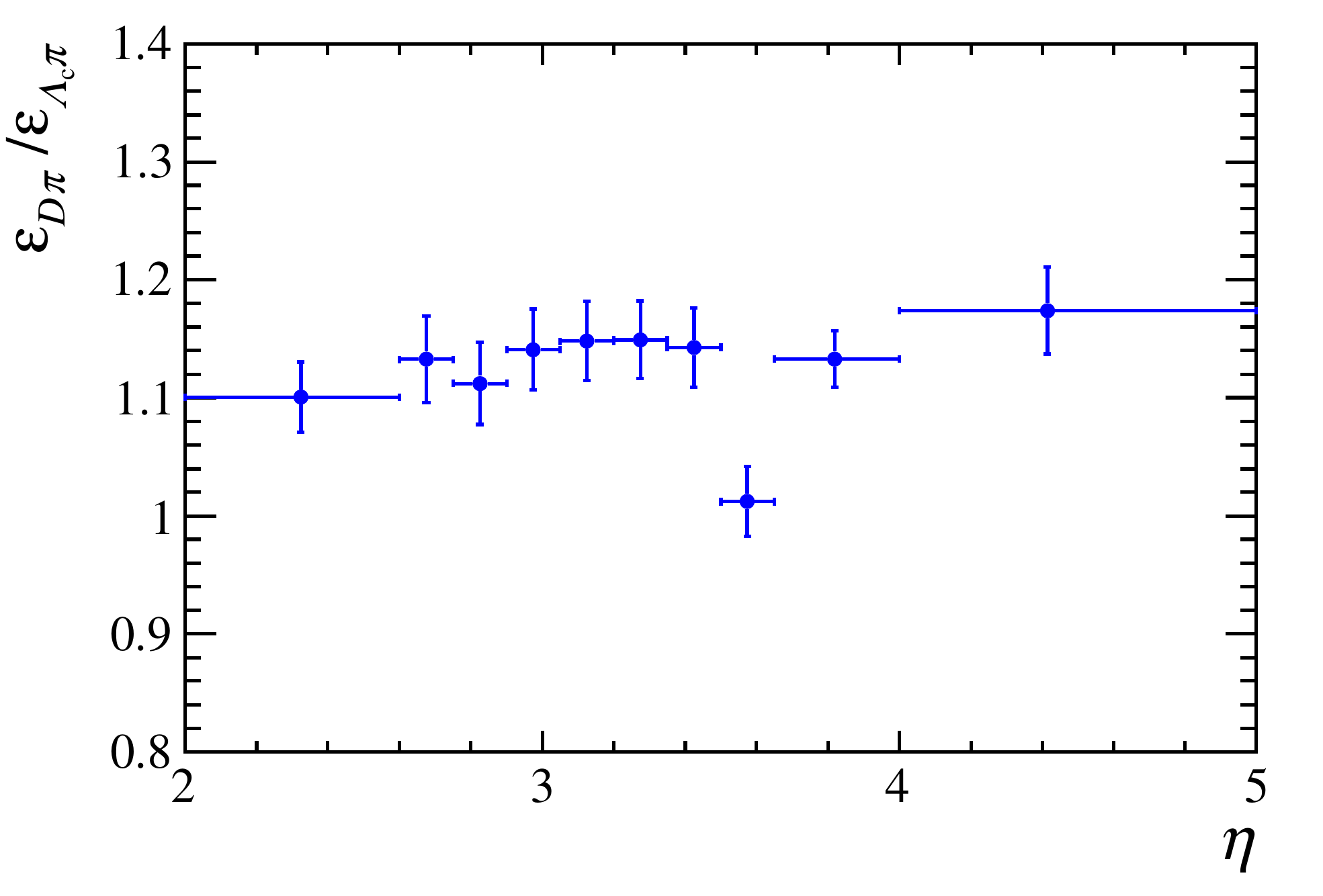}}
     \put(50,130){(a) LHCb}
     \put(280,130){(b) LHCb} 
  \end{picture} 
  \caption{\label{fig:effratio} Ratio of total selection efficiencies in bins of the (a) \pt and 
    (b) $\eta$ of the $b$ hadron. The horizontal error bars indicate the range of each bin in \pt or $\eta$ respectively.}
\end{figure}

The ratio of total selection efficiencies, $\varepsilon_{\BdDPi}/\varepsilon_{\LbLcPi}$, 
is shown in Fig.~\ref{fig:effratio}.
Fluctuations are included in the calculation of the efficiency-corrected 
yield ratio.

\section{Event yields}
\label{sec:Yields}

The dependences of \flfd on the \pt and
$\eta$ of the $b$~hadron are studied in the ranges $1.5 < \pt < 40$ \gevc and
$2 < \eta < 5$.  The event sample is sub-divided in 20 bins in \pt and 10
bins in $\eta$, with bin boundaries chosen to obtain approximately equal numbers
of \BdDPi candidates per bin.
The bin centres are obtained from simulated events without any selection applied,
and are defined as the  mean of the average \pt or $\eta$ of the \LbLcPi and 
\BdDPi samples in each bin.

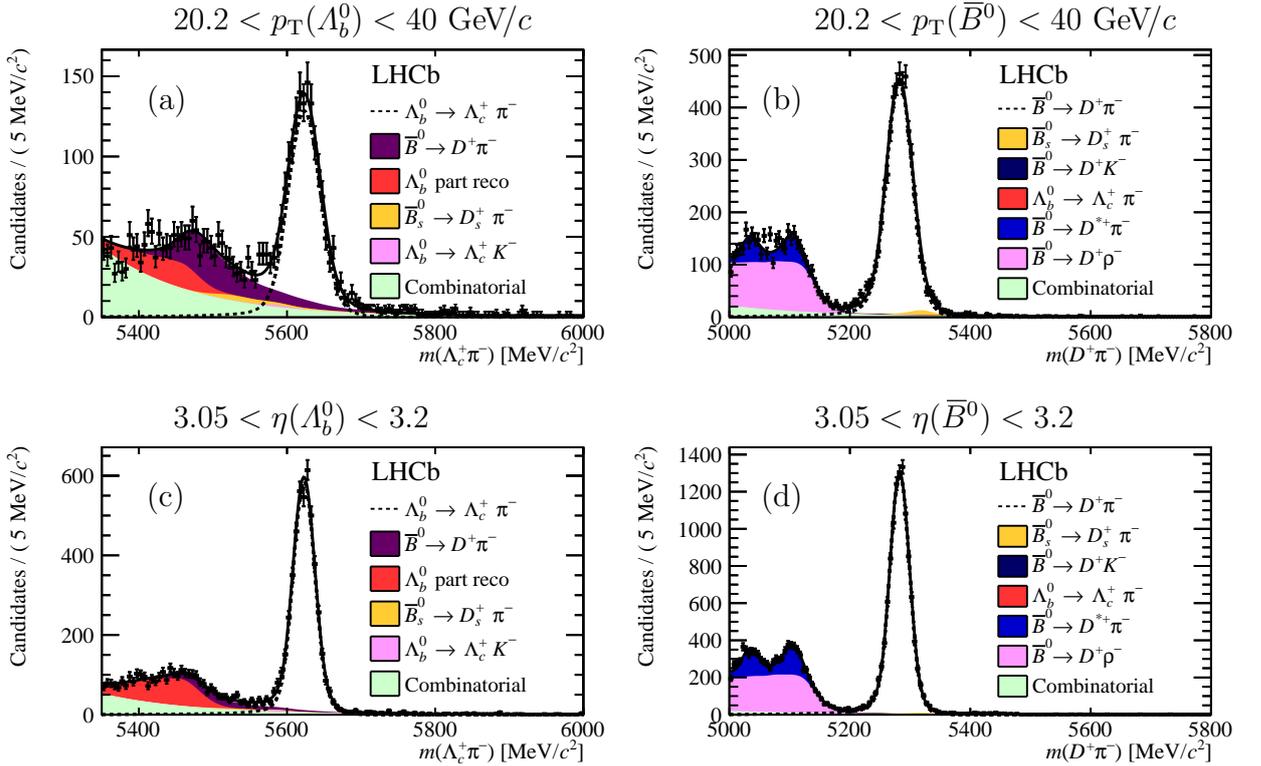
\begin{figure}[!b]
    \begin{picture}(500,300)(0,0)
     \put(-5,151){\includegraphics[bb=0 0 500 310, scale=0.48, clip=]{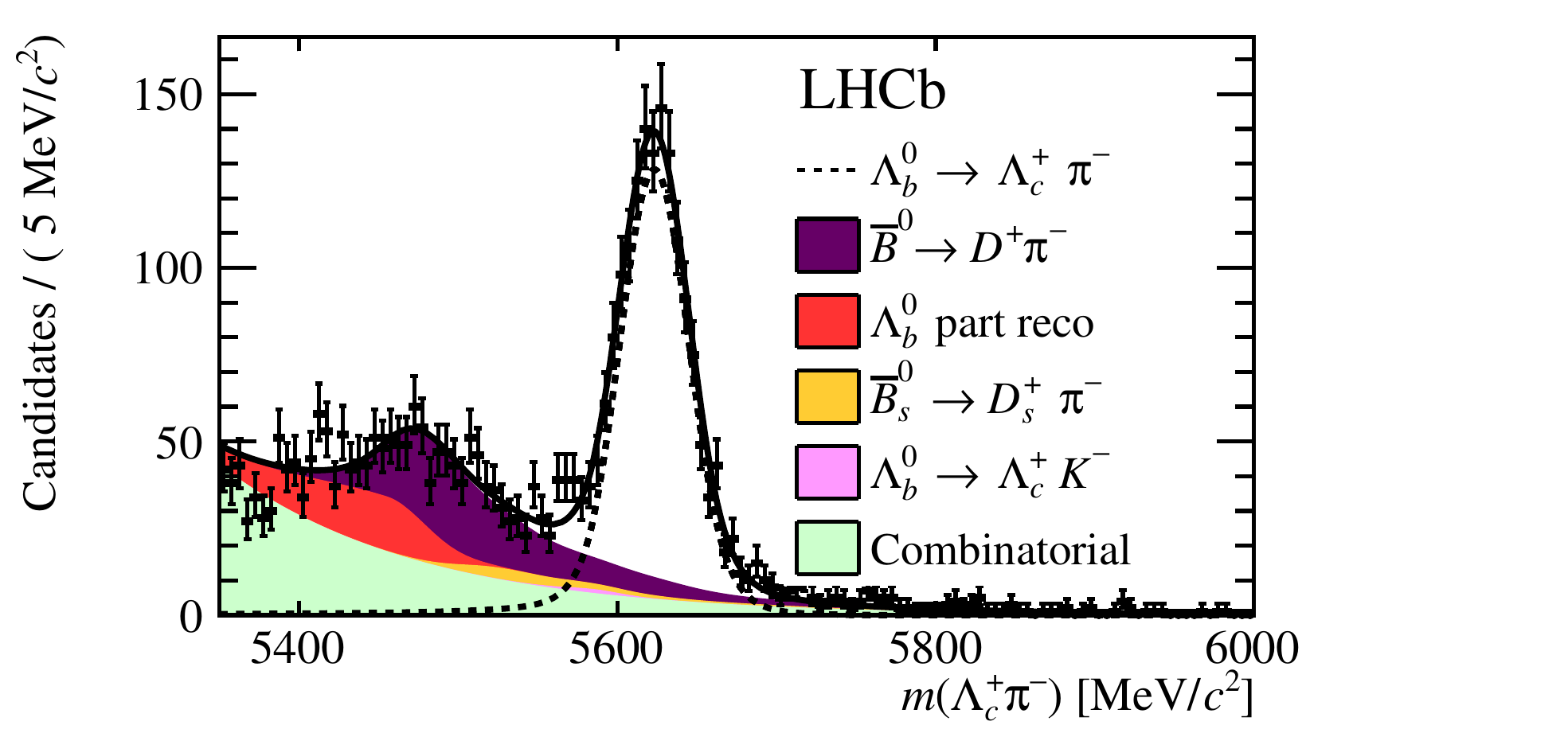}}
     \put(230,151){\includegraphics[bb=0 0 500 310, scale=0.48, clip=]{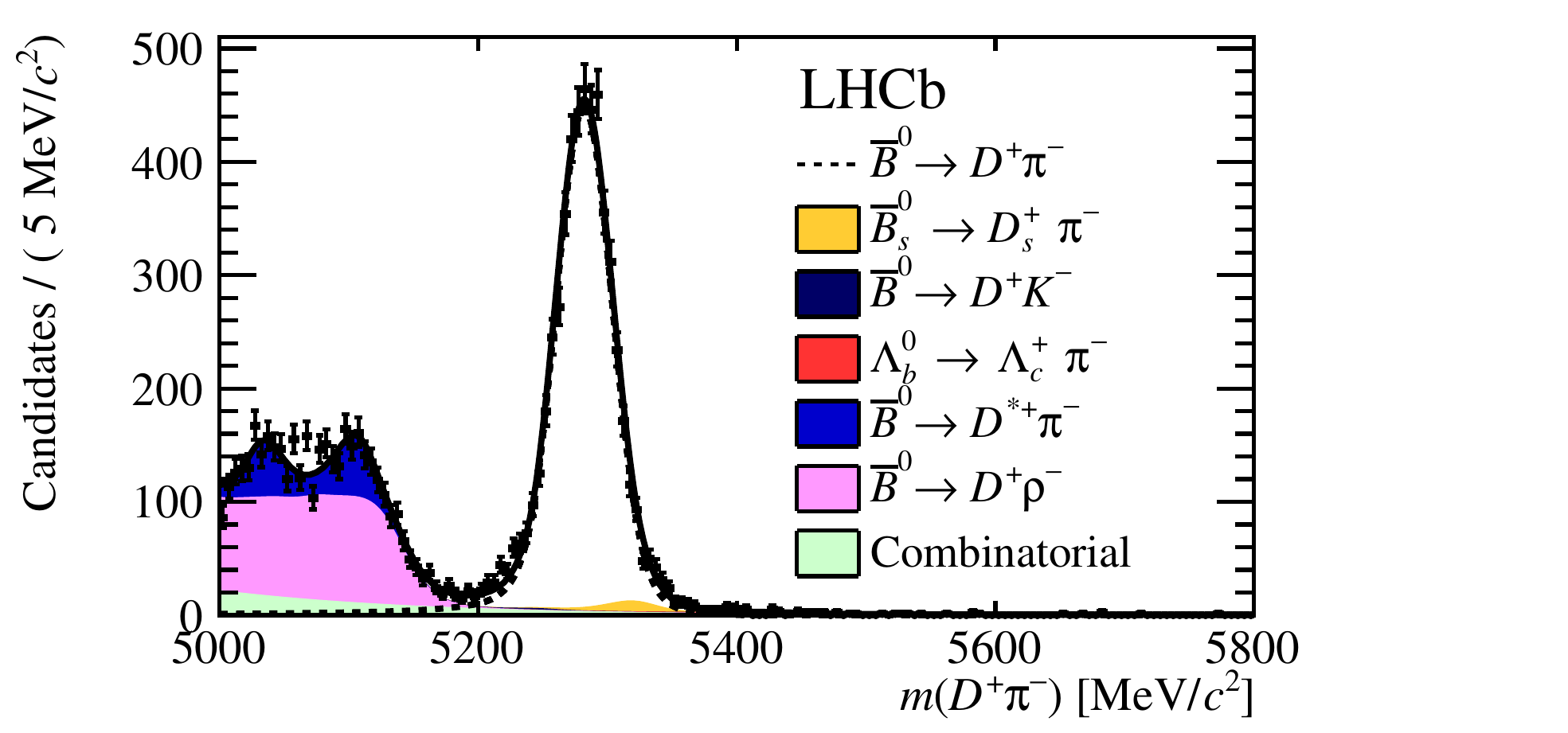}}
     \put(50,250){(a)}
     \put(60,280){$20.2<\pt(\Lb)<40\,\gevc$}
     \put(280,250){(b)} 
     \put(300,280){$20.2<\pt(\Bdb)<40\,\gevc$}

     \put(-5,1){\includegraphics[bb=0 0 500 310, scale=0.48, clip=]{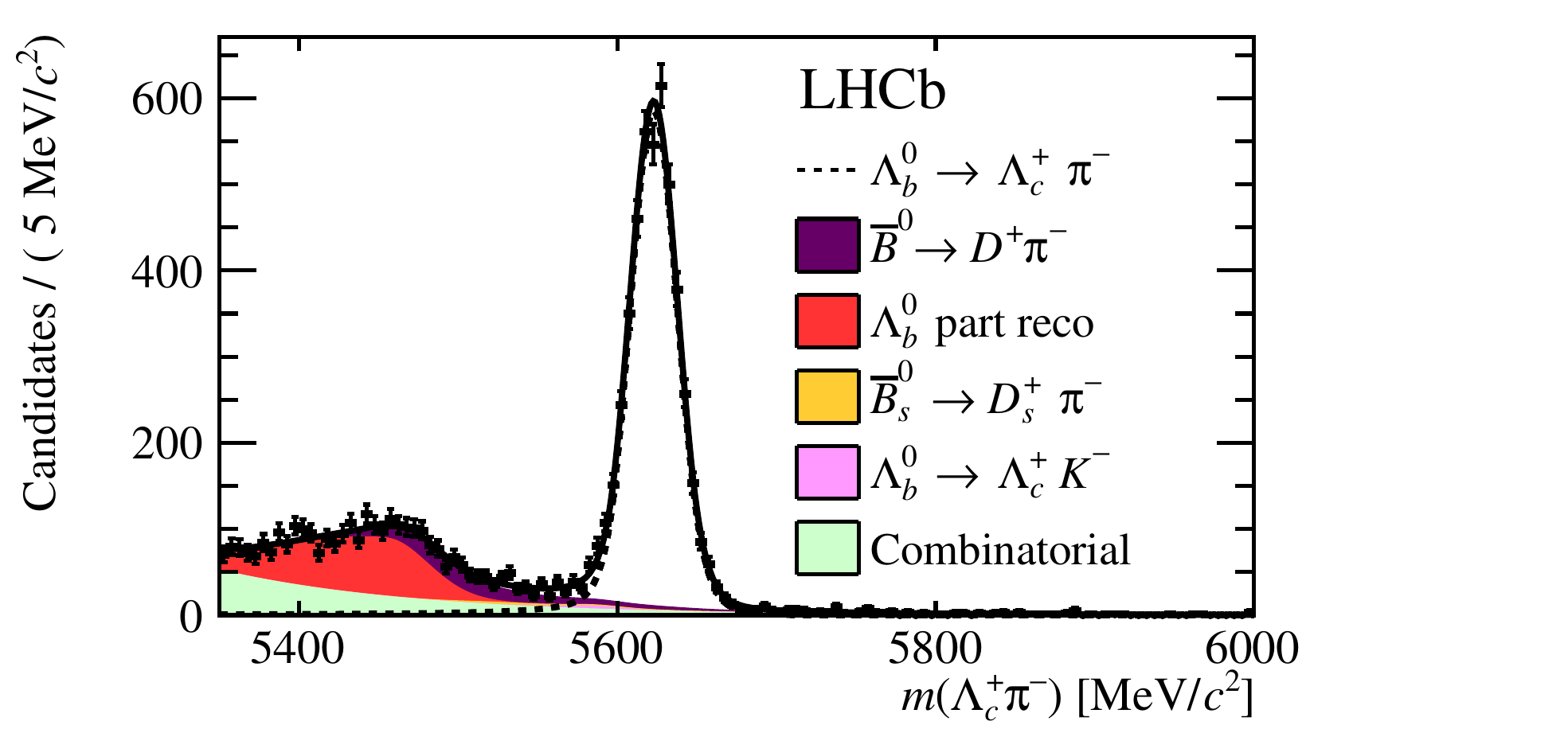}}
     \put(230,1){\includegraphics[bb=0 0 500 310, scale=0.48, clip=]{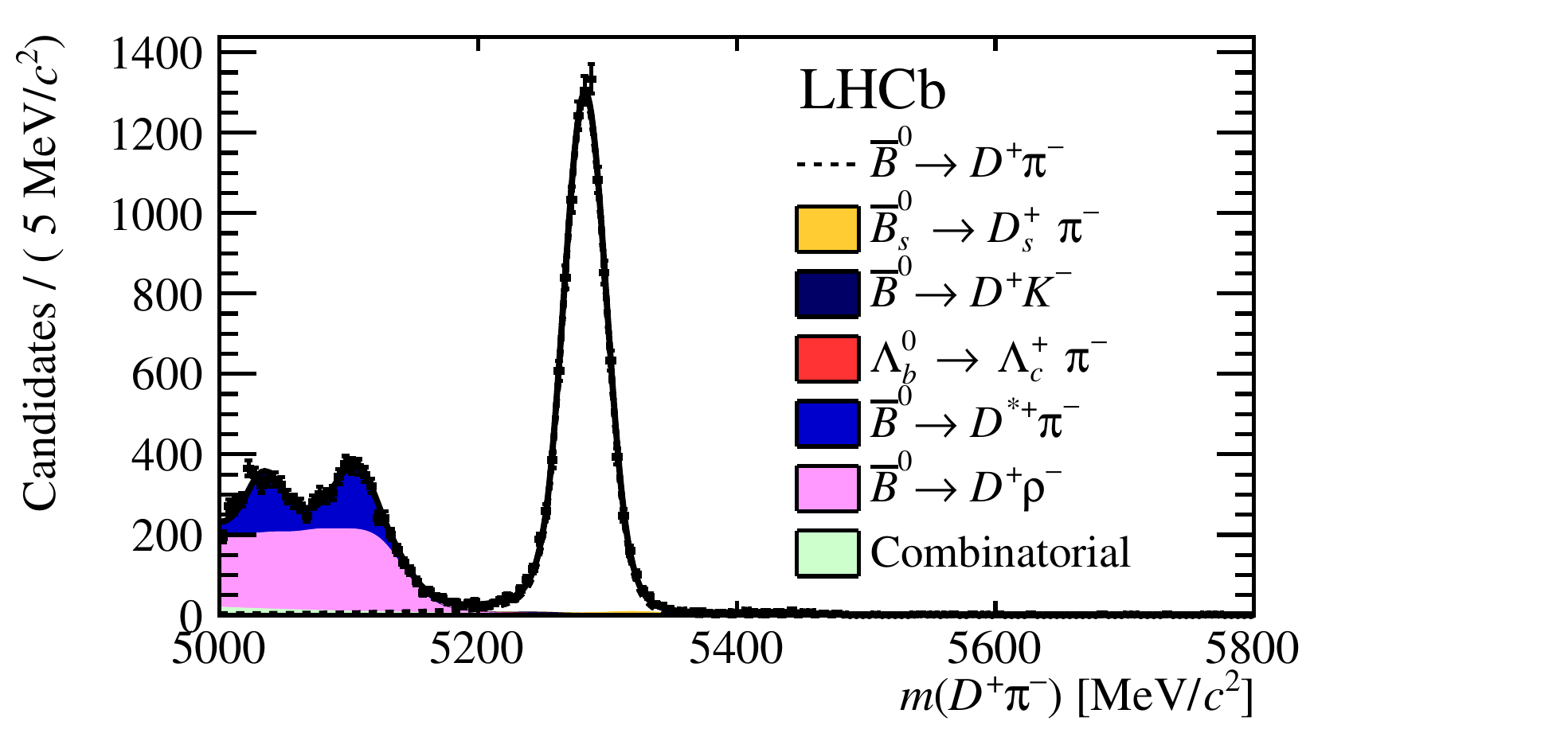}}
     \put(50,100){(c)}
     \put(60,130){$3.05<\eta(\Lb)<3.2$}
     \put(280,100){(d)} 
     \put(300,130){$3.05<\eta(\Bdb)<3.2$}
  \end{picture}
  \caption{\label{fig:masspeak} Invariant mass distributions of (a,c) $\Lc\pim$ candidates
    and (b,d) $\Dp\pim$ candidates for specific ranges in \pt and $\eta$ of the $b$~hadron,
    with fit projections overlaid.
    The different components are defined in the legend, where ``part reco" refers to the sum of partially reconstructed
    decays.}
\end{figure}

The yields of the two decay modes are determined from extended maximum
likelihood fits to the unbinned mass distributions of the reconstructed $b$-hadron
candidates, in each bin of \pt or $\eta$.  
To improve the mass resolution, the value of the beauty hadron mass is refit 
with the invariant mass of the charm hadron constrained to its known value~\cite{PDG2012}.
Example fits in the \pt bin with the lowest fitted signal yield and in an arbitrarily chosen $\eta$ bin are shown in
Fig.~\ref{fig:masspeak} for $\Lc\pim$ and $D^+\pim$ candidates. The total signal yields, 
obtained from fits to the total event samples, are $44\,859 \pm 229$ for the \LbLcPi sample
and $106\,197 \pm 344$ for the \BdDPi sample.

The signal mass shape is described by a modified Gaussian distribution
with power-law tails on either side to model the radiative tail and non-Gaussian detector
effects. 
The parameters of
the tails are obtained from simulated events and fixed in the fit. The mean and the width of the Gaussian distribution
are allowed to vary. 
 
Three classes of background are considered: partially reconstructed decays
with or without misidentified tracks, fully reconstructed
decays where at least one track is misidentified, and combinatorial background.  The shapes
of the invariant mass distributions for the partially reconstructed decays are
obtained using large samples of simulated events.  
For the \BdDPi sample, the decays \BdDRho and
\BdDstPi are modelled with non-parametric distributions~\cite{Cranmer:2000du}. 
The main sources for the \LbLcPi sample are  \LbLcRho and \LbSgPi decays, which are 
modelled with a bifurcated Gaussian function.
All these processes involve a neutral pion that is not included
in the candidate's reconstruction. 
 
The invariant mass distributions of the misidentified decays are affected by the
PID criteria. The shapes are obtained from simulated events, 
reweighted
according to the momentum-dependent particle identification
efficiency, with the mass hypothesis of the signal applied.  
The \BdDPi background in the \LbLcPi sample is most abundant in the
highest \pt bins, since the proton PID criteria are least effective in this kinematic region.

The Cabibbo-suppressed decays \LbLcK and \BdDK contribute to the background in the
\LbLcPi and \BdDPi fits, respectively, when the kaon of the $b$-hadron decay is misidentified as
a pion.  The yields of these backgrounds relative to the signal yield are
constrained in the fits, using \lhcb measurements of the relevant ratios of
branching fractions~\cite{LHCb-PAPER-2012-037,LHCb-PAPER-2013-056} and the
misidentification probabilities with their associated uncertainties.
 
The combinatorial background consists of events with random pions, kaons and
protons forming a mis-reconstructed \Dp or \Lc candidate, as well as genuine \Dp or \Lc
hadrons, that combine with a random pion.  The combinatorial background is
modelled with an exponential shape.  The slope is fixed in
the fit in each kinematic bin to the value found from a fit to the total sample.

\section{Results}
\label{sec:Results}

The study of the dependences of \flfd on the \pt and $\eta$ of the $b$ hadron and the 
measurement of the branching fraction of \LbLcPi decays are performed using candidates restricted to
the fiducial region $1.5 < \pt < 40$ \gevc and $ 2 < \eta < 5$. A discussion on the
systematic uncertainties related to these measurements can be found in the next section.

\begin{figure}[!h]
    \begin{picture}(500,180)(0,0)
     \put(-5,1){\includegraphics[scale=0.4]{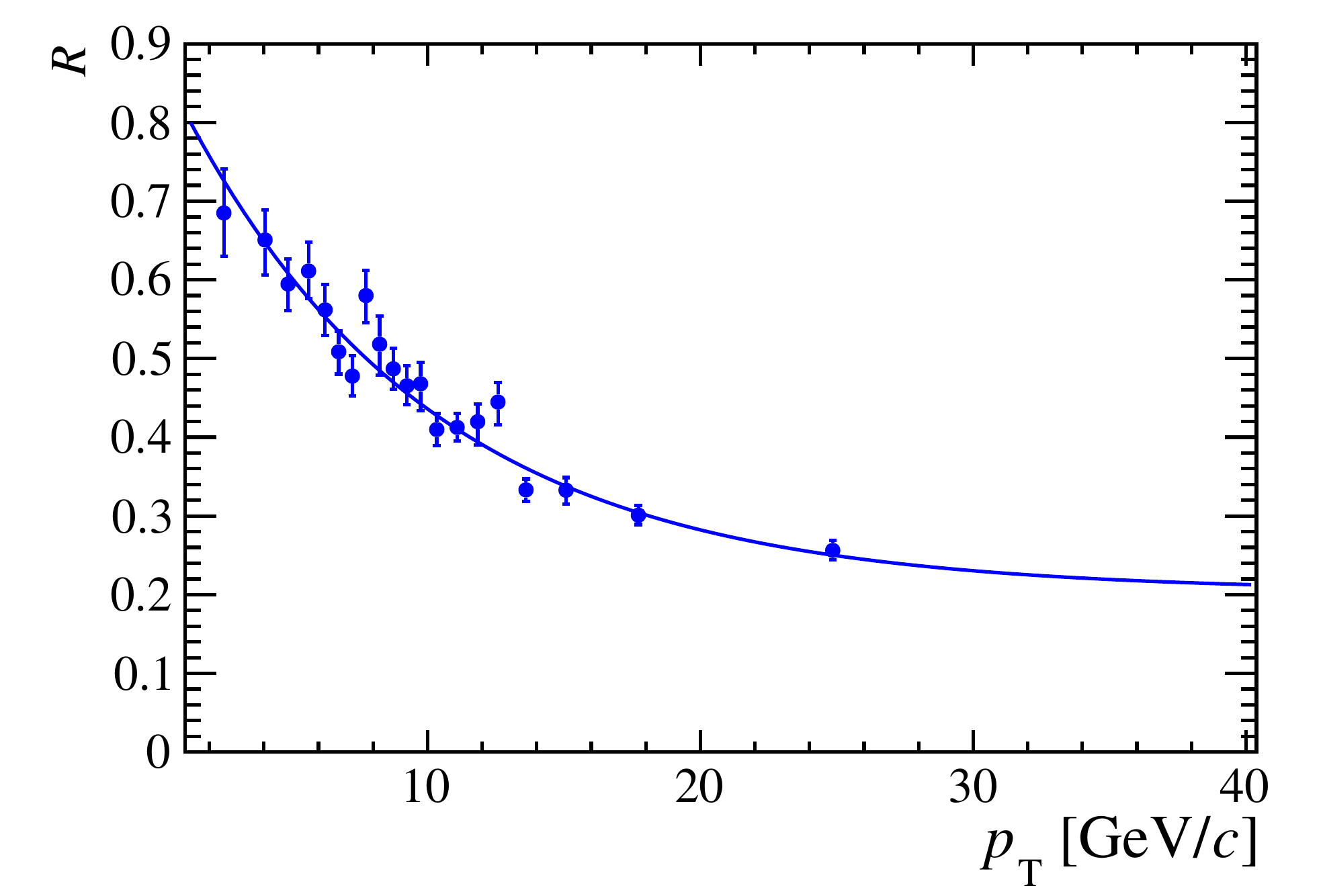}}
     \put(230,1){\includegraphics[scale=0.4]{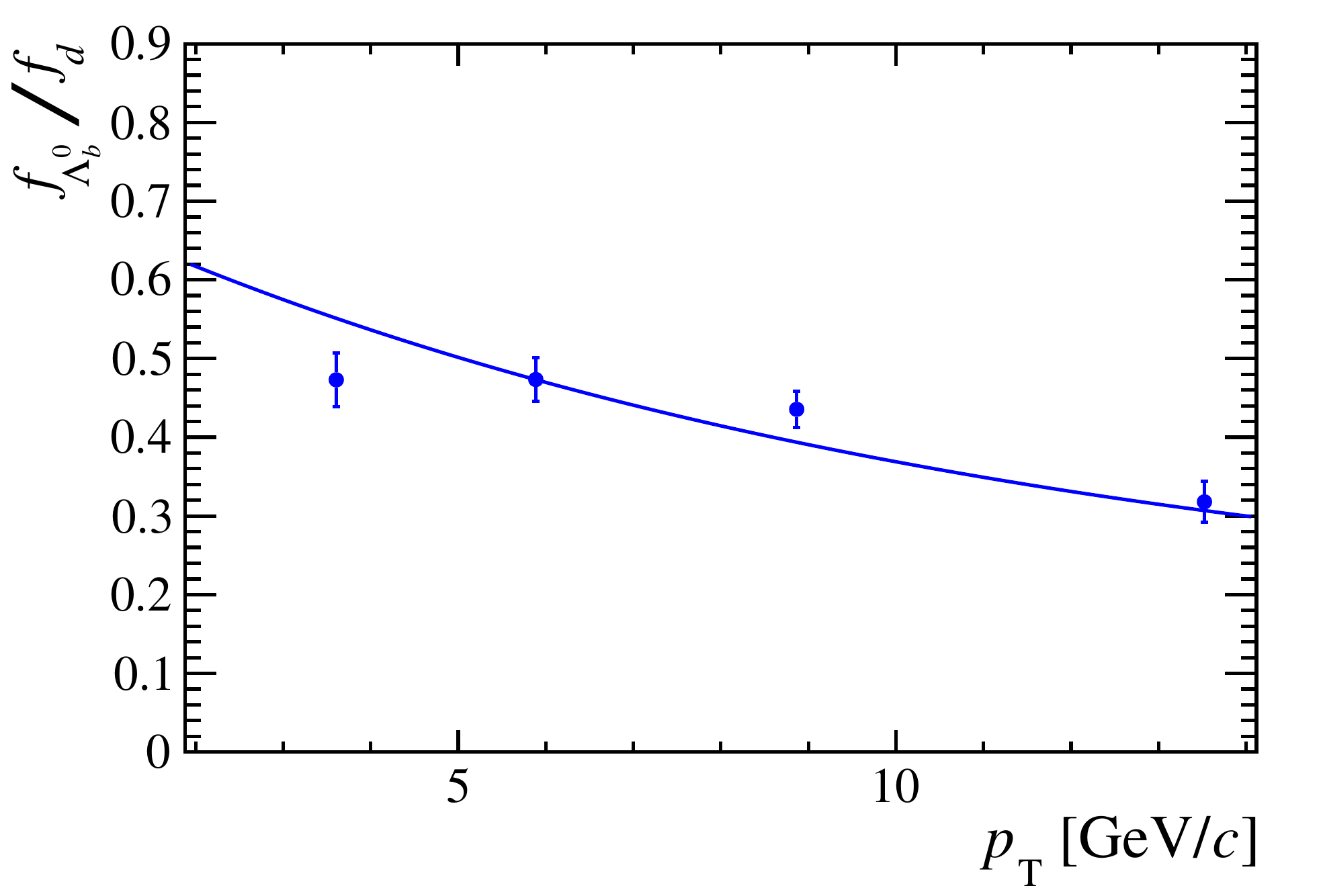}}
     \put(50,130){(a) LHCb}
     \put(280,130){(b) LHCb} 
  \end{picture} 
  \caption{\label{fig:flbfd-semi} (a) 
    Dependence of the efficiency-corrected ratio of yields, $\mathcal{R}$,
    between \LbLcPi and \BdDPi decays on the \pt of the beauty hadron, fitted with an exponential function. 
    The error bars on the data show the statistical and systematic uncertainties added in quadrature.
    (b) The resulting parametrisation is then fitted to the rescaled \flfd measurements from the semileptonic 
    analysis~\cite{LHCb-PAPER-2011-018}, to obtain the scale factor $\mathcal{S}$. The error bars include
    only the statistical uncertainty.}
\end{figure}

The ratio of efficiency-corrected event yields as a function of \pt is shown in
Fig.~\ref{fig:flbfd-semi}(a), and is fitted with an exponential function,

\begin{equation}
 \label{eq:ptRatio_syst} 
 \mathcal{R}(\pt) = a + \exp\left(b + c \times \pt [\gevc] \right),
\end{equation}
with
\begin{eqnarray}
\label{eq:ptRatioNumbers_syst}
a & = & +0.181 \pm 0.018 \pm 0.026, \nonumber \\
b & = & -0.391 \pm 0.023 ^{\,\, + 0.069} _{\,\, - 0.067}, \nonumber \\
c & = & -0.095 \pm 0.007 \pm 0.014 \,\,\, [\gevc]^{-1}, \nonumber
\end{eqnarray}
where the first uncertainties are statistical and the second systematic. 
The correlation matrix of the parameters is 
\begin{displaymath}
  \rho(a,b,c) = 
  \left(
  \begin{array}{ccc}
    1      & -0.22 & -0.94 \\
    -0.22 & 1      & -0.10 \\
    -0.94 & -0.10 & 1  
  \end{array}
 \right).
\end{displaymath}
The correlation between the parameters leads to a relatively large apparent uncertainty on the individual parameters.
Systematic uncertainties are not included in this matrix.
The \chisqndf value of the fit is 23.3/17, which corresponds to a p-value of
0.14.  

The $\eta$ dependence is described by a linear function,
\begin{equation}
\label{eq:etaRatio_syst}
\mathcal{R}(\eta) = a + b \times \left( \eta-\overline{\eta} \right),
\end{equation}
with
\begin{eqnarray}
\label{eq:etaRatioNumbers_syst}
a & = & 0.464 \pm 0.003 ^{\,\, + 0.008} _{\,\, - 0.010}, \nonumber \\
b & = & 0.081 \pm 0.005 ^{\,\, + 0.013}  _{\,\, - 0.009}, \nonumber \\
\overline{\eta} & = & 3.198, \nonumber
\end{eqnarray}
where the first uncertainties are statistical and the second systematic. 
The offset $\overline{\eta}$ is fixed to the average value of the measured $\eta$ distribution.
The correlation between the two fit parameters is negligible for this choice of $\overline{\eta}$.
The \chisqndf value of the fit is 13.1/8, corresponding to a p-value of 0.11.

To extract the scale factor $\mathcal{S}$ given in Eq.~\eqref{eq:DefnOfS}, the normalisation of $\mathcal{R}(x)$,
with fixed parameters $a$, $b$ and $c$, is allowed to vary in a fit to the published \flfd data~\cite{LHCb-PAPER-2011-018},
as shown in Fig.~\ref{fig:flbfd-semi}(b).
The result quoted in Ref.~\cite{LHCb-PAPER-2011-018} was measured as
a function of the \pt of the $\Lc\mu^{-}$ system.
A shift, estimated from simulation, is applied to the \pt values
to obtain the corresponding average \pt of the $b$~hadron for each bin. 
Furthermore, the semileptonic results are updated using recent determinations of
$\BR(\LcpKpi) = (6.84 \pm 0.24 ^{\,\,+0.21}_{\,\,-0.27})$\%~\cite{Zupanc:2013iki} and the ratio of lifetimes 
$(\tau_{B^+}+\tau_{B^0})/2\tau_{\Lb}=1.071 \pm 0.008$~\cite{LHCb-PAPER-2013-065,LHCb-PAPER-2014-003}.

The following value of the scale factor $\mathcal{S}$ is determined,
\begin{equation} \nonumber
  \label{eq:ScaleFactorValue}
  \mathcal{S} = 0.834 
  \overbrace{\pm 0.006 \, (\mathrm{stat})  ^{\,\,+0.023}_{\,\,-0.021} \, (\mathrm{syst})}^{\mathrm{hadronic}} 
  \overbrace{\pm 0.027 \, (\mathrm{stat})  ^{\,\,+0.058}_{\,\,-0.062} \, (\mathrm{syst})}^{\mathrm{semileptonic}},
\end{equation}
where the statistical and systematic uncertainties associated with the hadronic
and semileptonic measurement are shown separately.  
The \chisqndf value of the fit is 8.68/3, which corresponds to a p-value of 0.03.

\begin{figure}[!t]
    \begin{picture}(500,180)(0,0)
     \put(-5,1){\includegraphics[scale=0.4]{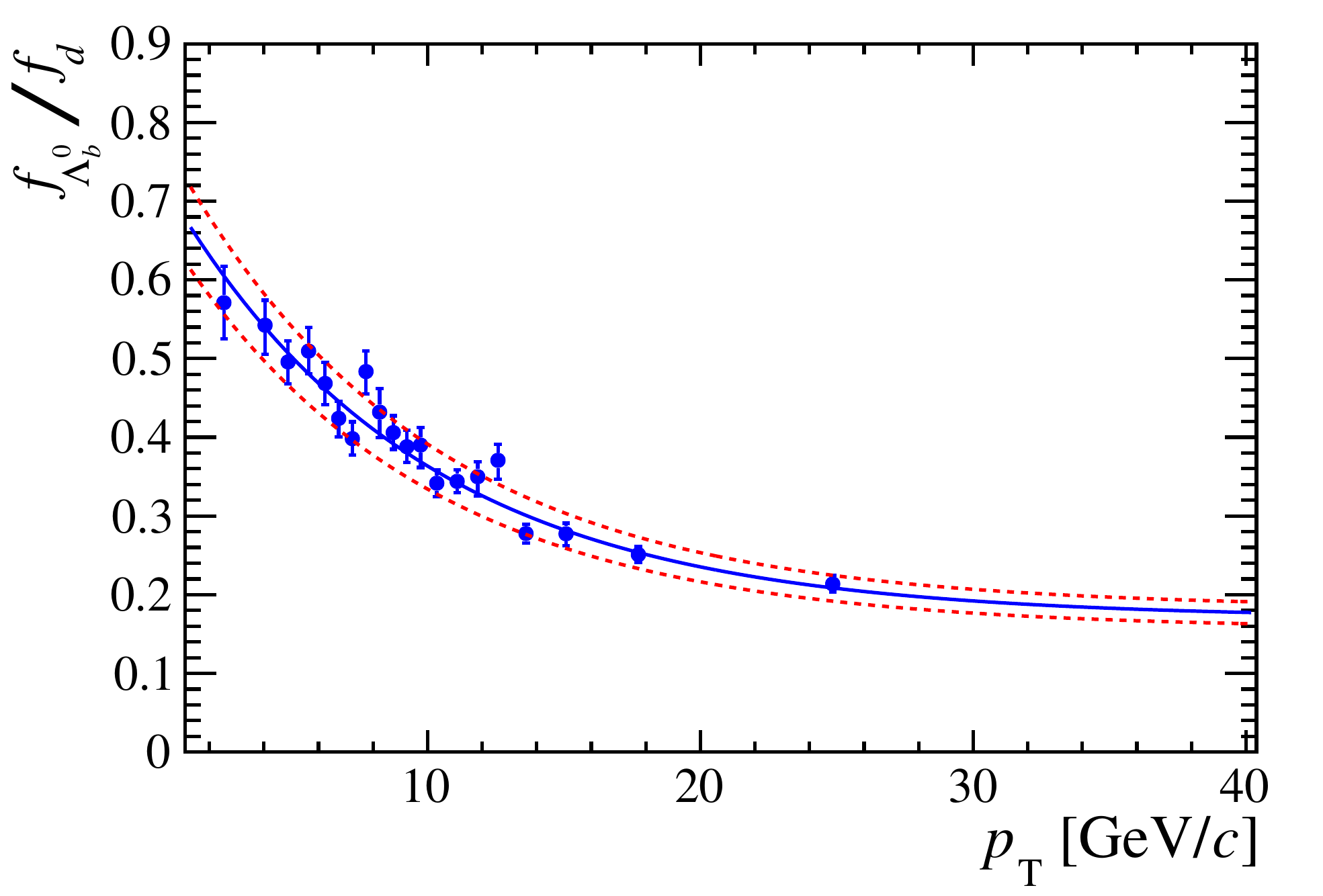}}
     \put(230,1){\includegraphics[scale=0.4]{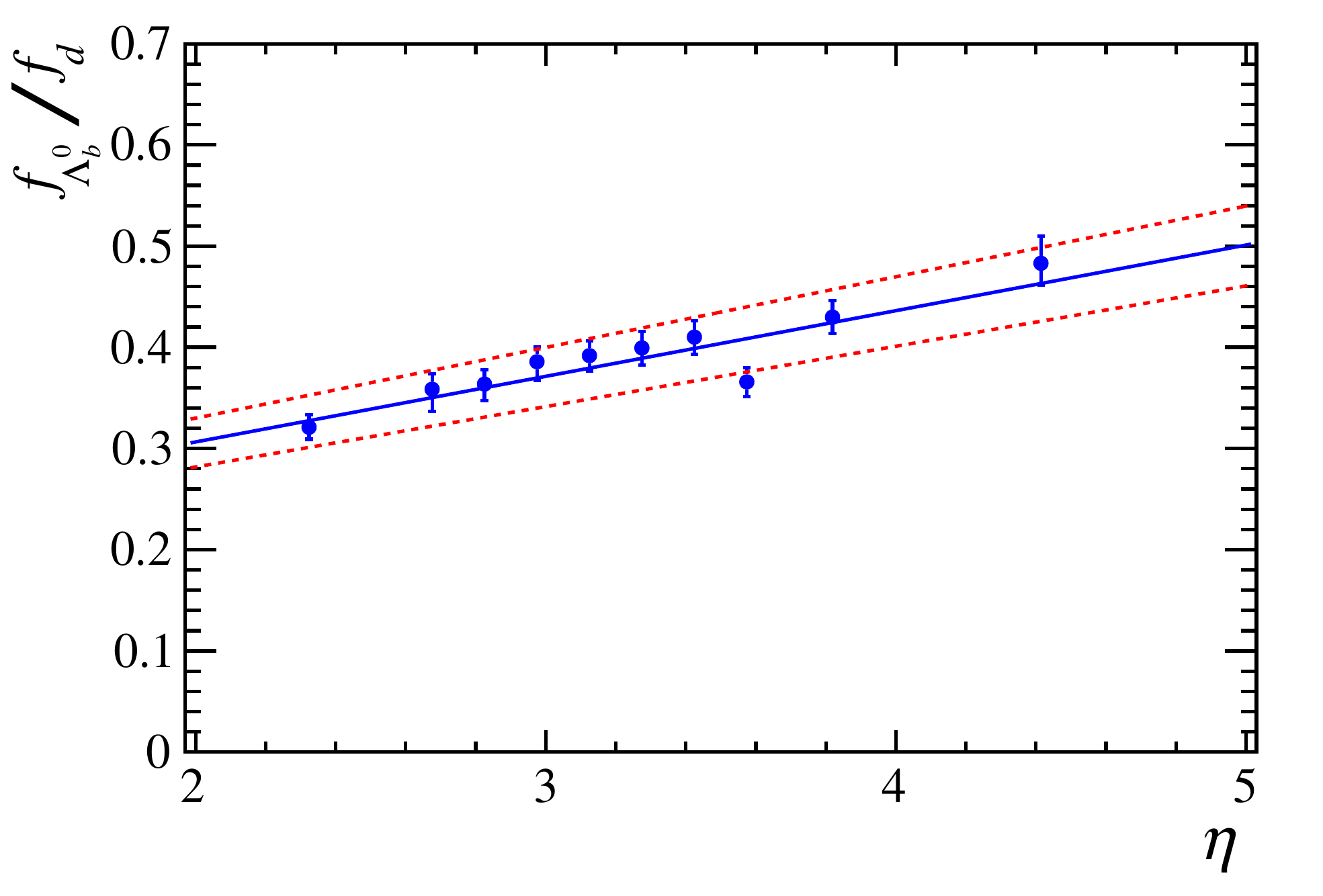}}
     \put(50,130){(a) LHCb}
     \put(280,130){(b) LHCb} 
  \end{picture}
  \caption{\label{fig:flbfd} Dependence of \flfd on the (a) \pt and (b) $\eta$ of the beauty hadron.
  To obtain this figure, the ratio of efficiency-corrected event yields is scaled to the absolute value 
  of \flfd from the semileptonic analysis~\cite{LHCb-PAPER-2011-018}. The error bars include the 
 statistical and systematic uncertainties associated with the hadronic measurement. The dashed red lines indicate the uncertainty
  on the scale of \flfd from the semileptonic analysis.}
\end{figure}

By multiplying the ratio of the efficiency-corrected yields $\mathcal{R}$ with the scale factor
$\mathcal{S}$, the dependences of \flfd on \pt and $\eta$ are
obtained. 
The \pt dependence is described with the exponential function
\begin{equation}
 \label{eq:flfd_model_pt} 
 \flfd(\pt) = a' + \exp(b' + c' \times \pt [\gevc]),
\end{equation}
with
\begin{eqnarray}
\label{eq:flfd_model_pt_numbers}
a' & = & +0.151 \pm 0.016 ^{\,\,+ 0.024} _{\,\,- 0.025} , \nonumber \\
b' & = & -0.573 \pm 0.040 ^{\,\,+ 0.101} _{\,\,- 0.097}, \nonumber \\
c' & = & -0.095 \pm 0.007 \pm 0.014  \,\,\, [\gevc]^{-1}, \nonumber
\end{eqnarray}
where the first uncertainty is the combined statistical and the second is the combined systematic from the hadronic and semileptonic measurements. 
The correlations between the three fit parameters change due to the uncertainty on the scale factor $\mathcal{S}$.  The correlation matrix of the parameters is 
\begin{displaymath}
  \rho(a',b',c') = 
  \left(
  \begin{array}{ccc}
    1 & 0.55 & -0.73 \\
    0.55 & 1 & -0.03 \\
    -0.73 & -0.03 & 1
  \end{array}
 \right).
\end{displaymath}
The $\eta$ dependence is described by the linear function
\begin{equation}
\label{eq:flfd_model_eta}
\flfd(\eta) = a' + b' \times \left( \eta-\overline{\eta} \right),
\end{equation}
with
\begin{eqnarray}
\label{eq:flfd_model_eta_numbers}
a' & = & 0.387 \pm 0.013 ^{\,\,+ 0.028} _{\,\,- 0.030}, \nonumber \\
b' & = & 0.067 \pm 0.005 ^{\,\,+ 0.012}  _{\,\,- 0.009}, \nonumber 
\end{eqnarray}
where the first uncertainty is the combined statistical and the second is the combined systematic from the hadronic and semileptonic measurements. 
The dependences of \flfd on the \pt and $\eta$ of the
$b$ hadron are shown in Fig.~\ref{fig:flbfd}.

The absolute value for \BR(\LbLcPi) is obtained by substituting the results for $\mathcal{S}$
and $\BR(\BdDPi)=(2.68 \pm 0.13) \times 10^{-3}$\cite{PDG2012} into Eq.~\eqref{eq:DefnOfS}.
The value for \BR(\LcpKpi) is also used in the determination of \flfd using 
semileptonic decays and therefore cancels in the final result. 
The branching fraction for \LbLcPi is measured to be
\begin{equation*}
\label{eq:BFResult}
\BF(\LbLcPi) = \BRLbResultRound,
\end{equation*}
where the first uncertainty is statistical, the second is systematic, 
the third is from the previous \lhcb measurement
of \flfd, and the fourth is due to the knowledge of
$\BR(\Bdb~\rightarrow~\Dp\pi^-)$. 
This value is
in agreement with the current world average~\cite{PDG2012}.  It also agrees within
2.4 standard deviations with the recent \lhcb measurement using \LbLcpKsPi decays~\cite{LHCb-PAPER-2013-061}, 
taking into account the correlated uncertainty 
from the semileptonic value for \flfd (6.1\%).  Combining the two \lhcb
measurements, and using a consistent value for the lifetime ratio of 
$(\tau_{B^+}+\tau_{B^0})/2\tau_{\Lb}=1.071 \pm 0.008$, we obtain
$\BF(\LbLcPi) = (4.46 \pm 0.36)\times 10^{-3}$, where the uncertainty is the combined
statistical and systematic uncertainty of both measurements.

\section{Systematic uncertainties}
\label{sec:Systematics}

Systematic uncertainties on the measurement of the relative
efficiency-corrected event yields of the \LbLcPi and \BdDPi decay modes relate to the
fit models and to the efficiencies of the PID, BDT and trigger selections.  
The effect of each systematic uncertainty on the 
efficiency-corrected yield ratio is calculated separately for each bin of
$\pt$ or $\eta$.
The systematic uncertainties are considered to be correlated across the bins, 
unless mentioned otherwise.
The effect of the systematic uncertainties on
the model of the $\mathcal{R}(x)$ dependence and
the measurement of \BR(\LbLcPi)
are determined by refitting the data points when
the $\mathcal{R}$ value in each bin is
varied by its associated uncertainty.
The various sources of systematic uncertainty are discussed below and summarised in Table~\ref{t:syst}. 

\begin{table}[b]
\begin{center}
\caption{Relative systematic uncertainties for the measurements of $\mathcal{R}(x)$ (first five columns) and \BR(\LbLcPi) (last column). The uncertainties from the various sources are uncorrelated and added in quadrature to obtain the total uncertainty. Sample size 
refers to the size of the simulated events sample.}
\label{t:syst}
{\small 
\begin{tabular}{l|rrr|rr|rrr} 
                                                         & \multicolumn{3}{c|}{$\pt$ bins}      & \multicolumn{2}{c|}{$\eta$ bins}  &  \\
                                                         & \multicolumn{3}{c|}{$\mathcal{R} = a + \exp(b+c \times \pt)$}   & \multicolumn{2}{c|}{$\mathcal{R} = a + b \times (\eta - \overline{\eta})$} & \multicolumn{3}{c}{\BR(\LbLcPi)} \\
                                                         & \multicolumn{1}{c}{$a$}     & \multicolumn{1}{c}{$b$}    & \multicolumn{1}{c|}{$c$}   & \multicolumn{1}{c}{$a$}      & \multicolumn{1}{c|}{$b$}   & &  &  \\    
\hline
\multicolumn{1}{l}{Fit model }                            &   \multicolumn{8}{c}{ }  \\  
\hline 
Signal                                   &   $^{+0.7}_{-0.4}$\% &   $^{+0.5}_{-0.2}$\% & $^{+0.2}_{-0.3}$\%  & $^{+0.3}_{-0.1}$\% &   $^{+1.1}_{-1.8}$\% &  &  $^{+0.2}_{-0.1}$\% & \\[0.3em]
Background                               &   $^{+5.5}_{-1.7}$\% &   $^{+2.8}_{-2.1}$\% & $^{+2.6}_{-1.1}$\%  & $^{+0.6}_{-0.1}$\% &   $^{+2.4}_{-4.7}$\% &  &  $^{+0.6}_{-0.0}$\% & \\
\hline
\multicolumn{1}{l}{Efficiencies}                           &   \multicolumn{8}{c}{ }  \\  
\hline
PID                                          &   0.0\% & 0.5\% & 2.5\%                          &   $-1.3$\% & 12.7\%  &  & $-1.1$\% & \\   
BDT                                            &   $^{+5.8}_{-7.6}$\% &   $^{-15.1}_{+14.2}$\%  &   $^{+~9.6}_{-10.2}$\%       &   $^{+1.3}_{-1.3}$\% &   $^{+4.7}_{-4.8}$\% &  &  $^{+2.3}_{-2.2}$\% & \\
Sample size                                     & $\pm$12.1\% & $\pm$9.0\% & $\pm$10.8\%       & $\pm$0.9\% & $\pm$9.3\% & & $\pm$1.2\%  & \\ 
Trigger                                     &   0.9\% & 1.0\% & 1.0\%                           &   $-0.3$\% & $-0.1$\%  & &  $-0.3$\% & \\   
\hline
\multicolumn{1}{l}{Other}                            &   \multicolumn{8}{c}{ }  \\  
\hline
Bin centre                                               & $\pm$0.3\% & $\pm$0.3\% & $\pm$0.1\%         & $\pm$0.1\% & $\pm$1.3\% & & 0.0\% & \\ 
\specialrule{.1em}{.05em}{.05em}
\specialrule{.1em}{.05em}{.05em}
\bf{Total} \normalfont                  & $^{+14.6}_{-14.5}$\% & $^{+17.1}_{-17.7}$\% & $^{+14.9}_{-14.9}$\%&  $^{+1.8}_{-2.1}$\% & $^{+16.6}_{-11.6}$\%  &  & $^{+2.6}_{-2.8}$\% & \\
\end{tabular}
}
\end{center}
\end{table}

The uncertainty due to the modelling of the signal shape is estimated by
replacing the modified Gaussian with two modified Gaussians, which share
the same mean but are allowed to have different widths.
In addition, the
parameters that describe the tails are varied by $\pm 10\%$ relative to their nominal
values, which is the maximum variation found for these parameters when leaving them
free in the fit. This affects the ratio of yields by a maximum of 0.3\%.

A possible variation of the slope of the combinatorial background shape  
across the bins is observed in a data sample of \LcPiWS candidates.
To account for this, the slope is varied from $\pm50\%$ in
the lowest \pt or $\eta$ bin to $\mp50\%$ in the highest \pt or $\eta$ bin.
The signal yield ratio varies by less than 1\%, with
the exception of one \pt bin which shows a variation of approximately 2\%.

The uncertainty on the shapes of partially reconstructed backgrounds is estimated by modelling 
them with a non-parametric distribution~\cite{Cranmer:2000du} for \LbSgPi and \LbLcRho decays and with two 
modified Gaussian distributions with tails on either side
for the \BdDstPi shape. The effect on the signal yield ratio is below 0.5\% in most bins,
increasing to about 2\% for the highest \pt bin.

The contribution of $b$-hadron decays without an intermediate $c$ hadron
is ignored in the fit. To evaluate the systematic uncertainty due to these decays, the $b$-hadron
mass spectra for candidates in the sidebands of the $c$-hadron mass distribution are examined. 
A contribution of $0.4\%$ relative to the signal yield is found in the \BdDPi decay mode, 
and its full size is taken as systematic uncertainty.
No contribution is seen in the \LbLcPi decay mode and no systematic uncertainty is assigned. 

The uncertainty on the PID efficiency and misidentification rate is estimated
by comparing the PID performance measured using simulated $\Dstar$ and $\PLambda$
calibration samples with that observed in simulated signal events.
The efficiency ratio varies by between 1\% and 4\% across the bins.

As discussed in Sec.~\ref{sec:EvtSel}, the simulated events are reweighted so that the distributions of 
quantities related to the track quality match the distributions observed in data.
The systematic uncertainty on the selection efficiency is obtained by recalculating the efficiency without this reweighting. 
The yield ratio varies by between 0.2\% and 6\%. 
In addition, there is a 5\% statistical uncertainty per bin due to the simulated sample size, 
which is uncorrelated across bins.

The uncertainty due to the trigger efficiency, caused by possible differences in
the response to a proton compared to a charged pion in the calorimeter, is
estimated to be about 0.4\%, taking into account that at most 10\% of the
events containing \LbLcPi candidates are triggered by the proton.
The systematic uncertainty due to the choice of bin centre is evaluated by redefining the bin centres 
using the average \pt or $\eta$ of the \Lb or \Bd sample only, instead of the mean of the \Lb and \Bd samples.

\section{Conclusions}
\label{sec:Conclusions}

The dependences of the production rate of \Lb baryons with respect to \Bd mesons are measured as functions of 
the transverse momentum \pt and of the pseudorapidity $\eta$ of the $b$~hadron.  
The \pt dependence is accurately described by an exponential function. 
The ratio of fragmentation fractions \flfd decreases by a factor of three in the range $1.5 < \pt < 40$~\gevc.  
The ratio of fragmentation fractions \flfd versus $\eta$ is described by a linear dependence in the range $2 < \eta < 5$.

The absolute scale of \flfd is fixed using the measurement of \flfd from semileptonic $b$-hadron 
decays~\cite{LHCb-PAPER-2011-018}.
The branching fraction of the decay \LbLcPi is determined with a total precision of 8\%, 
\begin{equation*}
\label{eq:BFResult_conc}
\BF(\LbLcPi) = \BRLbResultRound,
\end{equation*}
which is the most precise determination of a branching fraction of a beauty baryon to date.

\clearpage
\section*{Acknowledgements}
 
\noindent We express our gratitude to our colleagues in the CERN
accelerator departments for the excellent performance of the LHC. We
thank the technical and administrative staff at the LHCb
institutes. We acknowledge support from CERN and from the national
agencies: CAPES, CNPq, FAPERJ and FINEP (Brazil); NSFC (China);
CNRS/IN2P3 and Region Auvergne (France); BMBF, DFG, HGF and MPG
(Germany); SFI (Ireland); INFN (Italy); FOM and NWO (The Netherlands);
SCSR (Poland); MEN/IFA (Romania); MinES, Rosatom, RFBR and NRC
``Kurchatov Institute'' (Russia); MinECo, XuntaGal and GENCAT (Spain);
SNSF and SER (Switzerland); NASU (Ukraine); STFC and the Royal Society (United
Kingdom); NSF (USA). We also acknowledge the support received from EPLANET,
Marie Curie Actions and the ERC under FP7.
The Tier1 computing centres are supported by IN2P3 (France), KIT and BMBF (Germany),
INFN (Italy), NWO and SURF (The Netherlands), PIC (Spain), GridPP (United Kingdom).
We are indebted to the communities behind the multiple open source software packages on which we depend.
We are also thankful for the computing resources and the access to software R\&D tools provided by Yandex LLC (Russia).

\addcontentsline{toc}{section}{References}
\setboolean{inbibliography}{true}
\bibliographystyle{LHCb}
\bibliography{main,LHCb-PAPER,LHCb-CONF,LHCb-DP,Lb}


\end{document}